\documentclass[aps,prb,showpacs,superscriptaddress,twocolumn]{revtex4}
\usepackage{graphicx}
\begin{document}

\title{Magnetic and orbital ordering of RuO$_{2}$ planes in
  RuSr$_{2}$(Eu,Gd)Cu$_{2} $O$_{8}$}

\author{A. A. Aligia} \affiliation{Comisi\'{o}n Nacional de Energ\'{\i}a
  At\'{o}mica, Centro At\'{o}mico Bariloche and Instituto Balseiro, 8400
  S.C. de Bariloche, Argentina} \author{M. A.  Gusm\~{a}o}
\affiliation{Instituto de F\'{\i}sica, Universidade Federal do Rio Grande do
  Sul, CP 15051, 91501-970 Porto Alegre, Brazil}

\date{\today}

\begin{abstract}
  We start from an effective Hamiltonian for Ru ions in a square lattice,
  which includes the on-site interactions between $t_{2g}$ orbitals derived
  from Coulomb repulsion, and a tetragonal crystal-field splitting.  Using
  perturbation theory in the hopping terms, we derive effective Hamiltonians
  to describe the RuO$_{2}$ planes of RuSr$_{2}$(Eu,Gd)Cu$_{2}$O$_{8}$. For
  undoped planes (formal valence Ru$^{+5}$), depending on the parameters we
  find three possible orderings of spin and orbitals, and construct a phase
  diagram. This allows to put constraints on the parameters based on
  experimental data. When electron doping consistent with the hole doping of
  the superconducting RuO$_{2}$ planes is included, we obtain (for
  reasonable parameters) a double-exchange model with infinite
  antiferromagnetic coupling between itinerant electrons and localized
  spins. This model is equivalent to one used before [H. Aliaga and A. A.
  Aligia, Physica B {\bf 320}, 34 (2002)] which consistently explains the
  seemingly contradictory magnetic properties of
  RuSr$_{2}$(Eu,Gd)Cu$_{2}$O$_{8}$.
\end{abstract}


\maketitle

\section{Introduction}

In recent years, there has been much interest in ruthenates because of
their interesting magnetic and superconducting properties. For
example, SrRuO$_{3}$ is a ferromagnet that orders at $T_{M}=165$
K,\cite{long} Sr$_{2}$RuO$_{4}$ is an exotic $p$-wave superconductor
with transition temperature $T_{S}=1.5$ K,\cite{ishi} and
Sr$_{3}$Ru$_{2}$O$_{7}$ presents a metamagnetic transition\cite{perr}
and non-Fermi liquid behavior.\cite{grig} A close relationship between
ferromagnetic (FM) exchange and triplet $p$-wave superconductivity is
expected in analogy with $^{3}$He (Ref.\ \onlinecite{rice}) or from
bosonization studies in one dimension.\cite{bos,japa}

RuSr$_{2}$(Eu,Gd)Cu$_{2}$O$_{8}$ has a magnetic transition at $T_{M}
\sim 133$ K, and a superconducting transition at $T_{S}\sim 33$ K for
Eu or $T_{S}\sim 15-40$ K for Gd (depending on the conditions of
preparation and annealing).\cite{tall,bern} Below $T_{S}$,
superconductivity coexists with magnetic order, which was first
believed to be FM,\cite{bern,fain,pick,bute} since the magnetization
shows a rapid increase with magnetic field for fields below 5 T, and
the inverse magnetic susceptibility at high temperatures yields a
positive Curie constant $\Theta =100\pm 3$ K.\cite{bute} However,
neutron diffraction experiments found superlattice reflections
consistent with an usual antiferromagnetic (AF) order with
nearest-neighbor spins antiparallel in all three
directions.\cite{lynn} This seems difficult to reconcile with the
above mentioned magnetic properties, in particular with a positive
Curie constant. Nevertheless, a double-exchange model could
consistently account for these observations.\cite{alia}

The crystal structure of RuSr$_{2}$(Eu,Gd)Cu$_{2}$O$_{8}$ is similar to that
of YBa$_{2}$Cu$_{3}$O$_{7}$, except that Y is replaced by Eu or Gd, and the
CuO chain layer is replaced by a square planar RuO$_{2}$ layer, with
resulting tetragonal symmetry, except for small distortions typical of
perovskites.\cite{bern} The sequence of layers perpendicular to the
tetragonal axis is RuO$_{2}$/SrO/CuO$_{2}$/(Eu or Gd)/CuO$_{2}$/SrO. Several
experiments, like muon spin rotation,\cite{bern}
magnetization,\cite{bern,bute} electron paramagnetic resonance, and
ferromagnetic resonance,\cite {fain} demonstrate that the development of
superconductivity does nor affect the magnetic order. This suggests that, at
least as a first approximation, the superconducting CuO$_{2}$ planes and the
magnetic RuO$_{2}$ planes behave as separate entities related only by charge
transfer, as it happens with CuO planes and chains in
YBa$_{2}$Cu$_{3}$O$_{6+x}$.\cite{garc} Band structure calculations are
consistent with this picture.\cite{pick} From what is known for
YBa$_{2}$Cu$_{3}$O$_{6+x}$, \cite{garc,neum} a superconducting critical
temperature 
$T_{S}\sim$ 30--40 K suggests a doping of slightly less than 0.1
holes per CuO$_{2}$ plane. This implies a doping of $ \lesssim 0.2$
electrons to the RuO$_{2}$ planes with respect to the formal oxidation
states Ru$^{+5}$ and O$^{-2}$. Taking into account a certain degree of Ru-O
covalence, this doping is consistent with x-ray absorption near-edge
structure (XANES) experiments which suggest a Ru valence near 4.6.\cite{liu}
This situation is at variance with the compounds of the Ruddlesden-Popper
series (Cu,Sr)$_{n+1}$Ru$_{n}$O$_{3n+1}$ (like those mentioned above) for
which the formal oxidation state of Ru is Ru$^{+4} $.

The main features of the puzzling magnetic behavior of RuO$_{2}$
planes in RuSr$_{2}$(Eu,Gd)Cu$_{2}$O$_{8}$ were explained in terms of
a double-exchange model in which Ru$^{+5}$ spins have a strong Hund
coupling with a band of itinerant electrons.\cite{alia} Within this
picture, the undoped system presents usual AF ordering. Additional
electrons form FM polarons that tend to align easily in the direction
of an applied magnetic field.  Consequently, in spite of the AF order,
the magnetic susceptibility at temperatures $T>T_{M}$ can be well
described by $\chi =C/(T-\Theta )$, with $\Theta >0$, in agreement
with experiment.\cite{bute} Further support to the double-exchange
model is brought by the negative magnetoresistance above $T_{M}$, or
below $T_{M}$ for high enough magnetic field.\cite{awan,chen}
While this model has been successful in explaining several properties
of manganites,\cite{yuno,alia2} where the itinerant electrons are in
3d $e_{g}$ orbitals, there is so far no justification for its
application to ruthenates, where the relevant orbitals are the 4d
$t_{2g}$ ones,\cite{pick,mack,ogu} in which case crystal-field effects
are expected to be more important, and correlations should be smaller
due to the larger extent of the Ru 4d orbitals in comparison with Mn
3d ones.

In this work, we start from an effective Hamiltonian $H$ for the Ru 4d
$t_{2g}$ orbitals in a square lattice, after integrating out the O orbitals.
$H$ includes all atomic Coulomb interactions, and a tetragonal crystal-field
splitting $\Delta $. We treat $H$ in perturbation theory in the effective
Ru-Ru hopping. Since the effective parameters are difficult to estimate, the
quantitative validity of this strong-coupling approach is difficult to
address. However, for reasonable parameters our results are consistent with
experiment, and confirm results previously obtained with the double-exchange
model.  In the case of Sr$_{2}$RuO$_{4}$, which has been studied in more
detail, there is a wide range of proposed parameters, but it is clear that
the correlations are significant, and the system is believed to be in the
intermediate-coupling regime.\cite{wern,lieb} Notice that approaches that
neglect quantum fluctuations should assume smaller interactions to avoid
magnetic ordering in Sr$_{2}$RuO$_{4}$, while instead RuO$_{2}$ planes in
RuSr$_{2}$(Eu,Gd)Cu$_{2}$O$_{8}$ do order magnetically at $T_{M}$.

The paper is organized as follows. In Section II we present the model and
discuss its parameters. In Section III we describe the eigenstates an
energies of the local Hamiltonian. In Section IV we derive effective
Hamiltonians that describe spin and orbital degrees of freedom in the
undoped case, after integrating out the charge fluctuations. Section V
contains the phase diagram for this case. In Section VI we discuss the
effective Hamiltonians for the doped case, and their relation to the double
exchange model. Our results are summarized and discussed in Section VII.

\section{The model}

We start with an effective model for the 4d $t_{2g}$ orbitals of Ru ions in
a square lattice. It can be derived from an appropriate multiband model for
RuO$_{2}$ planes by a canonical transformation eliminating Ru-O hopping
terms \cite{pet} or by the cell perturbation method if Ru-O covalence were
important.\cite{jeff,pet2} The Hamiltonian is
\begin{equation} \label{eq:Hparts}
H=\sum_{i}(H_{I}^{i}+H_{CF}^{i})+H_{h}\;,
\end{equation}
where $H_{I}^{i}$ contains the local interaction terms at site $i$,
$H_{CF}^{i}$ is a tetragonal crystal-field splitting, and $H_{h}$ contains the
hopping terms which we restrict to nearest neighbors. Since $H_{I}^{i}$
contains only intra-site interactions, we assume for it the same form as for
an isolated Ru ion, neglecting spin-orbit coupling. This form can be
calculated in a straightforward way using known methods of atomic
physics.\cite {cond,ball} Expanding the Coulomb interaction term
$e^{2}/|{\bf r}_{1}-{\bf r}_{2}|$ in spherical harmonics, all Coulomb
integrals can be expressed in terms of Slater parameters $F_{0}$, $F_{2}$,
and $F_{4}$ (as done earlier\cite{bati} for $e_{g}$ orbitals). Here we write
$H_{I}^{i}$ using the Kanamori parameters (which seem to be more popular in
condensed matter \cite {kana,bocq}) for $t_{2g}$ orbitals: $U = F_{0} +
4F_{2} + 36F_{4}$, $J=J^{\prime } = 3F_{2} + 20 F_{4}$, and $U^{\prime} =
U-2J$. Then,
\begin{widetext}
\begin{equation} \label{eq:hi}
H_{I}^{i} = U\sum_{\alpha } n_{i\alpha \uparrow } n_{i\alpha
\downarrow } + \frac{1}{2} \sum_{\alpha \neq \beta, \sigma \sigma
^{\prime }} (U^{\prime } n_{i\alpha \sigma} n_{i\beta \sigma^{\prime
}} + Jd_{i\alpha \sigma }^{\dagger }d_{i\beta \sigma ^{\prime
}}^{\dagger }d_{i\alpha \sigma ^{\prime }}d_{i\beta \sigma })
+J^{\prime }\sum_{\alpha \neq \beta }d_{i\alpha \uparrow }^{\dagger
}d_{i\alpha \downarrow }^{\dagger }d_{i\beta \downarrow }d_{i\beta
\uparrow }\;,
\end{equation}
where $n_{i\alpha \sigma } = d_{i\alpha \sigma }^{\dagger } d_{i\alpha
  \sigma}$, and $d_{i\alpha \sigma }^{\dagger }$ creates an electron in the
$t_{2g}$ orbital $\alpha $ ($xy$, $yz$ or $zx$) with spin $\sigma $ at site
$i$. 

\mbox{}
\end{widetext}
Choosing $z$ as the tetragonal axis, we write the crystal field term in the
form
\begin{equation} \label{eq:hcf}
H_{CF}^{i}=\Delta (\sum_{\sigma } d_{ixy\sigma }^{\dagger } d_{ixy\sigma }
-1 ) \;,
\end{equation}
in such a way that it changes sign under an electron-hole transformation.

Denoting by $\delta =\pm {\bf \hat{x}}$, $\pm {\bf \hat{y}}$ the four
vectors that connect a site with its four nearest neighbors, the hopping
term has the form
\begin{eqnarray} \label{eq:hh}
H_{h} &=&-t\sum_{i\sigma }(d_{i+{\bf \hat{x}},zx\sigma }^{\dagger
}d_{izx\sigma }+d_{i+{\bf \hat{y}},yz\sigma }^{\dagger }d_{iyz\sigma } +
\text{H.c.})  \nonumber \\ 
&& \mbox{} - t^{\prime } \sum_{i\delta \sigma } d_{i+\delta ,xy\sigma
}^{\dagger} d_{ixy\sigma }\;.  
\end{eqnarray}
Notice that, since we neglect the distortions, electrons occupying $zx$
($yz$) orbitals do not hop in the $y$ ($x$) direction due to the symmetry of
the intermediate O 2p orbitals.\cite{pet,pet2}

While the parameters of $H$ are difficult to estimate, we expect that
the order of magnitude of $t$ and $t^{\prime }$ is near 1/4
eV\cite{alia,wern,lieb,pet,pet2} (see also Section VII).  Since the
exchange interactions are not expected to be strongly screened in the
solid, one may estimate $J$ from atomic spectra.\cite{moore} From the
low lying levels of Ru$^{+}$ (with three holes in the 4d shell), we
obtain $F_{2}\sim 863$ cm$^{-1}$ and $F_{4}\sim 78$ cm$^{-1}$, leading
to $J\sim 0.5$ eV. Optical experiments in Sr$_{2}$RuO$_{4}$ suggest
that $U\sim 1.5$ eV.\cite{yoko} Notice that the expectation value of
the Coulomb repulsion in any state with two electrons and total spin
$S=1$ should be positive. This implies, for two different $t_{2g}$
orbitals in the atomic case, $F_{0}-5F_{2}-24F_{4}>0$, or
\begin{equation} \label{eq:cond}
U-3J=U^{\prime }-J>0\;. 
\end{equation}
For a Slater determinant with both $e_{g}$ orbitals one obtains
$F_{0}-8F_{2}-9F_{4}>0$.\cite{bati} This condition is expected to be more
restrictive than Eq.~(\ref{eq:cond}), since $F_{2}$ is usually more than one
order of magnitude larger than $F_{4}$. For example, using the above
estimates for $F_{2}$ and $F_{4}$ this gives $F_{0}>0.94$ eV and $U^{\prime}
- J > 0.17$ eV. We assume Eq.~(\ref{eq:cond}) to be valid in general.
Otherwise for large $|\Delta|$ there is a charge-transfer instability of the
ground state for the undoped system. The physics of Sr$_{2}$RuO$_{4}$
suggests that $\Delta$ is small and negative.\cite{lieb}

\begin{widetext}

\begin{table}[t]
\caption{\label{tab:states}Eigenstates and energies of
  $H_{I}^{i}+H_{CF}^{i}$ for two and three 
particles. Here $u_{j},v_{j}>0$, $u_{j}^{2}+v_{j}^{2}=1$,
$u_{2}^{2}=[1-(\Delta -J^{\prime }/2)/r_{2}]/2$, $r_{2}=$ $[(\Delta 
-J^{\prime }/2)^{2} + 2(J^{\prime })^{2}]^{1/2}$, $u_{3}^{2}=[1-\Delta
/r_{3}]/2$, and $r_{3}=[\Delta^{2}+(J^{\prime})^{2}]^{1/2}$. A prime
indicates new appearance of the same irreducible representation of the point
group. States obtained by applying the spin lowering operator $S^-$ or
rotation of $\pi /2$ in the $xy$ plane are not shown.}
\begin{ruledtabular}
\begin{tabular}{lll}
notation & eigenstate & energy \medskip \\ \hline \vspace*{-10pt} \\
$|2a_{1}0\rangle $ & $[u_{2}d_{xy\downarrow }^{\dagger }d_{xy\uparrow
}^{\dagger }-\frac{1}{\sqrt{2}}v_{2}(d_{zx\downarrow }^{\dagger
}d_{zx\uparrow }^{\dagger }+d_{yz\downarrow }^{\dagger }d_{yz\uparrow
}^{\dagger })]|0\rangle $ & $U+J^{\prime }/2-r_{2}$ \medskip \\ 
$|2a_{1}^{\prime }0\rangle $ & $[u_{2}d_{xy\downarrow }^{\dagger
}d_{xy\uparrow }^{\dagger }+\frac{1}{\sqrt{2}}v_{2}(d_{zx\downarrow
}^{\dagger }d_{zx\uparrow }^{\dagger }+d_{yz\downarrow }^{\dagger
}d_{yz\uparrow }^{\dagger })]|0\rangle $ & $U+J^{\prime }/2+r_{2}$ \medskip \\ 
$|2b_{1}0\rangle $ & $\frac{1}{\sqrt{2}}(d_{zx\downarrow }^{\dagger
}d_{zx\uparrow }^{\dagger }-d_{yz\downarrow }^{\dagger }d_{yz\uparrow
}^{\dagger })|0\rangle $ & $U-J^{\prime }-\Delta $ \medskip \\ 
$|2b_{2}0\rangle $ & $\frac{1}{\sqrt{2}}(d_{yz\uparrow }^{\dagger
}d_{zx\downarrow }^{\dagger }-d_{yz\downarrow }^{\dagger }d_{zx\uparrow
}^{\dagger })|0\rangle $ & $U^{\prime }+J-\Delta $ \medskip \\ 
$|2x0\rangle $ & $\frac{1}{\sqrt{2}}(d_{xy\downarrow }^{\dagger
}d_{yz\uparrow }^{\dagger }-d_{xy\downarrow }^{\dagger }d_{yz\uparrow
}^{\dagger })|0\rangle $ & $U^{\prime }+J$\medskip  \\ 
$|2a_{2}11\rangle $ & $d_{yz\uparrow }^{\dagger }d_{zx\uparrow }^{\dagger
}|0\rangle $ & $U^{\prime }-J-\Delta $\medskip  \\ 
$|2x11\rangle $ & $d_{xy\uparrow }^{\dagger }d_{yz\uparrow }^{\dagger
}|0\rangle $ & $U^{\prime }-J$ \medskip \\ \hline \vspace*{-10pt} \\ 
$|3a_{1}\frac{1}{2}\frac{1}{2}\rangle $ & $\frac{1}{\sqrt{2}}d_{xy\uparrow
}^{\dagger }(d_{yz\uparrow }^{\dagger }d_{zx\downarrow }^{\dagger
}-d_{yz\downarrow }^{\dagger }d_{zx\uparrow }^{\dagger })|0\rangle $ & 
$3U^{\prime}$ \medskip \\ 
$|3a_{2}\frac{1}{2}\frac{1}{2}\rangle $ & $\frac{1}{\sqrt{2}}d_{xy\uparrow
}^{\dagger }(d_{zx\downarrow }^{\dagger }d_{zx\uparrow }^{\dagger
}-d_{yz\downarrow }^{\dagger }d_{yz\uparrow }^{\dagger })|0\rangle $ & $
U+2U^{\prime }-J-J^{\prime }$ \medskip \\ 
$|3b_{1}\frac{1}{2}\frac{1}{2}\rangle $ & $\frac{1}{\sqrt{6}}[d_{xy\uparrow
}^{\dagger }(d_{yz\uparrow }^{\dagger }d_{zx\downarrow }^{\dagger
}+d_{yz\downarrow }^{\dagger }d_{zx\uparrow }^{\dagger })-2d_{xy\downarrow
}^{\dagger }d_{yz\uparrow }^{\dagger }d_{zx\uparrow }^{\dagger }]|0\rangle $
& $3U^{\prime }$ \medskip \\ 
$|3b_{2}\frac{1}{2}\frac{1}{2}\rangle $ & $\frac{1}{\sqrt{2}}d_{xy\uparrow
}^{\dagger }(d_{zx\downarrow }^{\dagger }d_{zx\uparrow }^{\dagger
}+d_{yz\downarrow }^{\dagger }d_{yz\uparrow }^{\dagger })|0\rangle $ & $
U+2U^{\prime }$ \medskip \\ 
$|3x\frac{1}{2}\frac{1}{2}\rangle $ & $(u_{3}d_{xy\downarrow }^{\dagger
}d_{xy\uparrow }^{\dagger }-v_{3}d_{yz\downarrow }^{\dagger }d_{yz\uparrow
}^{\dagger })d_{zx\uparrow }^{\dagger }|0\rangle $ & $U+2U^{\prime }-J-r_{3}$
\medskip \\ 
$|3x^{\prime }\frac{1}{2}\frac{1}{2}\rangle $ & $(v_{3}d_{xy\downarrow
}^{\dagger }d_{xy\uparrow }^{\dagger }+u_{3}d_{yz\downarrow }^{\dagger
}d_{yz\uparrow }^{\dagger })d_{zx\uparrow }^{\dagger }|0\rangle $ & $
U+2U^{\prime }-J+r_{3}$ \medskip \\ 
$|3b_{1}\frac{3}{2}\frac{3}{2}\rangle $ & $d_{xy\uparrow }^{\dagger
}d_{yz\uparrow }^{\dagger }d_{zx\uparrow }^{\dagger }|0\rangle $ & $
3U^{\prime }-3J$
\end{tabular}
\end{ruledtabular}
\end{table}

\end{widetext}

\section{Eigenstates of the local Hamiltonian}

The local part $H_{I}^{i} + H_{CF}^{i}$ can be easily diagonalized. To
describe the undoped system, we need the eigenstates with three electrons,
and those with two and four electrons are needed when the effects of the
hopping term $H_{h}$ or doping are included. We denote the eigenstates by
$|in\Gamma SM\rangle $, where $i$ is the site index, $n$ is the number of
electrons, $\Gamma $ denotes the symmetry (irreducible representation of the
point group $D_{4h}$ or symmetry of the basis function for the
two-dimensional representation), $S $ is the total spin and $M$ its
projection on the tetragonal axis $z$. If $S=0$, $M$ is suppressed. For
simplicity, we drop the site index in this Section. The subscript $g$ is
also dropped in the irreducible representations. Some eigenstates and its
energies are listed in Table I. The remaining ones for $n=3$ and $n=2$ are
obtained by applying the operator $S^{-}$ or a rotation of $\pi /2$ around
$z$ to those listed. The corresponding results for $n=4$ can be obtained
from those of $n=2$ using electron-hole symmetry: replace creation by
annihilation operators with the opposite spin, $d_{i\alpha \sigma }^{\dagger
}\to d_{i\alpha ,-\sigma }$, replace the vacuum by the state with all
$t_{2g}$ orbitals occupied, change the sign of $\Delta$, and add
$U+4U^{\prime }-2J=5U^{\prime }$ to the resulting energies.

Since we started with a local interaction Hamiltonian $H_{I}^{i}$ with
full rotational symmetry, the symmetry group of $H_{I}^{i}+H_{CF}^{i}$ is
actually higher than $D_{4h}$. For example rotations around $z$ of the
orbitals $yz$ and $zx$ keeping the $xy$ fixed leave $H_{I}^{i}+H_{CF}^{i}$
invariant. As a consequence the state $|2b_{1}0\rangle $ is degenerate with 
$|2b_{2}0\rangle $, and the states $|3a_{1}\frac{1}{2}M\rangle $ and 
$|3a_{2}\frac{1}{2}M^{\prime }\rangle $ also become degenerate. This degeneracy is
broken if the conditions $U=U^{\prime }+2J$, $J^{\prime }=J$ are relaxed,
but a degeneracy between $|3a_{1}\frac{1}{2}M\rangle $ and $|3b_{1}
\frac{1}{2}M^{\prime }\rangle $ persists which is broken only 
if an exchange interaction 
$J$ between the orbitals $yz$ and $zx$ different from the other two 
is introduced.

For $n=3$, the ground state is the spin quadruplet
$|3b_{1}\frac{3}{2}M\rangle$ if $|\Delta |\leq \sqrt{15}J$, while for
$|\Delta |\geq \sqrt{15}J $ the ground state is also four-fold degenerate,
but it is the spin and orbital $E_{g}$ doublet $|3x\frac{1}{2}M\rangle $,
$|3y\frac{1}{2}M\rangle$. These two possibilities lead to two different
effective Hamiltonians in the undoped case, after integrating out the charge
degrees of freedom.

\section{Effective Hamiltonians for undoped planes}

In this Section we construct effective Hamiltonians $H_{\text{eff}}$ for the
undoped case, using second-order degenerate perturbation theory in $H_{h}$.
Depending on the ground state of $H-H_{h}$, there are two possibilities for
$H_{\text{eff}}$.

\subsection{$\Delta ^{2}<15J$}

In this case the ground-state manifold of $H-H_{h}$ is the spin quadruplet
$|i3b_{1} \frac{3}{2}M \rangle$ at each site $i$. The degeneracy is lifted
by second-order contributions in which the intermediate states have two
nearest-neighbor sites with 2 and 4 electrons, both with total spin $S=1$
and both with the same symmetry $B_{2g}$, $x$ or $y$. The different matrix
elements are easily calculated using Eq.~(\ref{eq:hh}) and Table I. We omit
the details. The resulting $H_{\text{eff}}$ is a Heisenberg model for the
effective $S=3/2$ spins:

\begin{equation}  \label{eq:hea}
H_{\text{eff}}^{a} = K \sum_{\langle ij\rangle } ({\bf S}_{i} {\bf .S}_{j} -
\frac{9}{4}) \;; \qquad K = \frac{4(t^{2} + t^{\prime 2})}{9(U+2J)}\;. 
\end{equation}
The coupling constant $K$ turns out to be independent of $\Delta $. The
ground state of this model is a two-sublattice antiferromagnet with
antiparallel nearest-neighboring spins. We call it AFI. The energy per site
can be calculated accurately enough using spin waves, and is given
by\cite{sw}
\begin{equation}  \label{eq:ea}
E_{AFI} - E_{3b_{1}3/2} \simeq -2 KS(S+0.158) = -4.974K, 
\end{equation}
where $E_{3b_{1}3/2}$ is the energy of the state $|i3b_{1}\frac{3}{2} M
\rangle $ given in Table I.

\subsection{ Large $\Delta ^{2}$}

For $\Delta ^{2} > 15 J$ the ground state of $H_{I}^{i}+H_{CF}^{i}$ is the
spin and orbital doublet $|i3\gamma \frac{1}{2}M\rangle $, with $\gamma =x$
or $y$, which we denote briefly as $|i\gamma \sigma \rangle $. The number of
intermediate states is much larger than in the previous case, and
$H_{\text{eff}}$ becomes very complicated. Since for $\Delta ^{2}>15J$ the
structure of the states involved in the derivation is already very similar
to that for $\Delta \rightarrow \pm \infty $ (as can be checked by 
inspection of Table I), we restrict the calculation to
this case. The result is
\begin{eqnarray}  \label{eq:heb}
H_{\text{eff}}^{b} &=&\sum_{i}\{J_{A}\sum_{\gamma }P_{i\gamma }P_{i+{\bf 
\hat{\gamma}},\gamma }({\bf S}_{i}{\bf .S}_{i+{\bf
\hat{\gamma}}}-\frac{1}{4})  \nonumber \\ 
&&-\sum_{\delta }P_{ix}P_{i+\delta ,y}(J_{F}{\bf S}_{i}{\bf .S}_{i+{\bf 
\delta }}+A)\}\;, 
\end{eqnarray}
where now ${\bf S}_{i}$ are spin-1/2 operators, $P_{i\gamma }$ are the
orbital projectors
\begin{equation} \label{eq:pr}
P_{i\gamma }=\sum_{\sigma }|i\gamma \sigma \rangle \langle i\gamma \sigma
|\;, 
\end{equation}
and
\begin{eqnarray}  \label{eq:ctes}
J_{A} &=&2t^{2}(\frac{1}{U^{\prime }+J}+\frac{1}{U^{\prime }+3J})\,, ~~
J_{F}=\frac{2Jt^{2}}{U'^{\,2}-J^{2}}\,,  \nonumber \\
A &=&\frac{t^{2}}{U^{\prime }-J}-\frac{J_{F}}{4}\;. 
\end{eqnarray}
The first term of $H_{\text{eff}}^{b}$ is a one-dimensional
interaction, FM in the orbital degrees of freedom, and AF in spin. The
term proportional to $A$ is a spin independent AF orbital interaction,
while the $J_{F}$ term is FM in spin and AF in orbital variables.

Clearly, there are two possible competing ground states of
$H_{\text{eff}}^{b}$: i) a ferromagnetic orbital ordering (all sites
$|ix\sigma \rangle $, for example), with spin degrees of freedom
determined by the critical one-dimensional AF Heisenberg model; ii) a
spin FM and orbital N\'eel ordered phase (for example $|ix\uparrow
\rangle $ in one sublattice and $|iy\uparrow \rangle $ in the
other). In the first case, for finite $\Delta$, a small AF coupling
$J_{\perp }$ between the chains appears [see Eqs.~(\ref{eq:heb2}) and
(\ref{eq:jperp})], which yields long-range order at $T=0$. We call
this phase AFII.  From Bethe ansatz results,\cite{bah} the energy of
this phase for $\Delta \rightarrow \pm \infty $ is
\begin{equation}  \label{eq:eaf}
E_{\text{AFII}} = E_{3x1/2}-J_{A}\ln 2\;. 
\end{equation}
The second phase will be denoted FM-AFO, and its ground-state
energy is
\begin{equation} \label{eq:ef}
E_{\text{FM-AFO}}-E_{3x1/2} =
-2(A+\frac{J_{F}}{4})=-\frac{2t^{2}}{U^{\prime }-J}\;.
\end{equation}

\section{The phase diagram}

We now turn to the construction of a phase diagram for undoped planes,
comparing the energies of the phases described in the previous
Section, but now for arbitrary $\Delta$. 
Since the correction terms for $\sqrt{15}J <|\Delta |<+\infty $ are
small, we do not expect any new phases to appear in this interval,
except perhaps near the borderline between two phases, as we will
discuss at the end of this section. 

\begin{figure}[t]
  \begin{center}
\includegraphics[width=8cm]{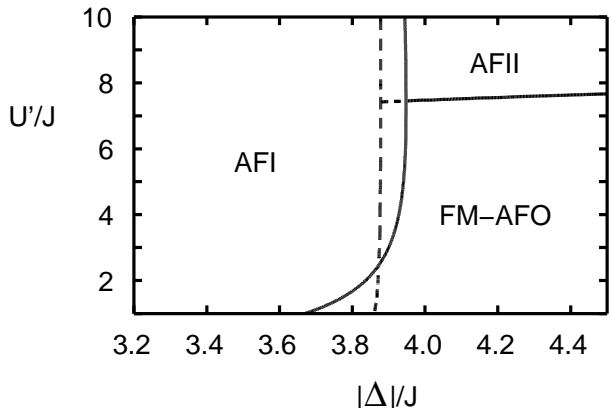}    
  \end{center}
  \caption{Phase diagram of undoped RuO$_{2}$ planes for $t = t^{\prime} = 
  J/4$ (thin line) and $t = t^{\prime} = J$ (thick line). The phases are 
  labeled as explained in the text.}
 \label{fig:res}
\end{figure}
 
The energy of the AFI phase is still given by Eq.~(\ref{eq:ea}), due
to the independence of $H^a_{\text{eff}}$ [Eq.~(\ref{eq:hea})] on the
crystal-field sppliting. The most important change occurs in the part
of $H^b_{\text{eff}}$ [Eq.~(\ref{eq:heb})] that describes the AFII
phase, in which an interchain coupling is generated. Thus, the
effective Hamiltonian of the AFII phase, assuming that the chains are
oriented along the $x$ direction, is
\begin{equation} \label{eq:heb2}
H_{\text{eff}}^{b2}=\sum_{i}\{J_{A}^{\prime }({\bf S}_{i}{\bf .S}_{i+{\bf 
\hat{x}}}-\frac{1}{4})+J_{\perp }{\bf S}_{i}{\bf .S}_{i+{\bf
\hat{y}}}+C\}\;.
\end{equation}
where $J_{A}^{\prime }$ is the AF coupling along the chains, which
coincides with $J_{A}$ for $|\Delta| \to \infty$, $J_{\perp
}=4(C_{\uparrow \uparrow }-C_{\uparrow \downarrow })$, $C=C_{\uparrow
\uparrow }+C_{\uparrow \downarrow }$, and $C_{\uparrow \uparrow }$
($C_{\uparrow \downarrow }$) is the negative correction of the energy
due to virtual hoppings from site $i$ to $i+{\bf \hat{y}}$ and back
when the spin of both sites are equal (opposite). The expressions for
$J_{A}^{\prime }$, $C_{\sigma \sigma ^{\prime }}$, as well as the
energy of the
FM-AFO phase are lengthy but straightforward
to obtain, and we do not reproduce them here. Instead, $J_{\perp }$
takes a simpler form,
\begin{widetext}
\begin{equation}  \label{eq:jperp}
J_{\perp }=\frac{2J^{4}(t^{2}+t^{\prime 2})}{\left( \Delta ^{2}+J^{2}\right)
\left[ u^{\prime }+2(r_{3}-J)\right] \left( U^{\prime 2}+4\left[ \Delta
^{2}+J^{2}+U^{\prime }(2r_{3}-J)\right] \right)} \;, 
\end{equation}
where $r_{3}=(\Delta ^{2}+J^{2})^{1/2}$. Eq.~(\ref{eq:cond}) and the
stability condition of the AFII phase against AFI for vanishing $t$ and
$t^{\prime }$ ($\Delta ^{2}>15J$) imply $J_{\perp }/J_{A}<6.80\times
10^{-4}$.

\mbox{}\end{widetext}

The energy of the AFII phase up to second order in $H_{h}$ is given by
\begin{equation} \label{eq:eaf2}
E_{\text{AFII}}^{\prime }=E_{3x1/2}-J_{A}^{\prime }\ln 2+C+E_{J_{\perp }}\;,
\end{equation}
where $E_{J_{\perp }}$ is the correction due to the interchain coupling.
This correction can be calculated treating the $J_{\perp}$ term in a
mean-field approximation, by a straightforward generalization of Schulz's
results\cite{schu} for the case in which each chain has $z$ nearest
neighboring chains (he considered $z=4$ while we have $z=2 $). The energy
gain due to the appearance of a spontaneous staggered expectation value of
the spin projection in the chain direction $m=(-1)^{i}\langle
S_{i}^{z}\rangle $ is (for any sign of $J_{\perp }$)
\begin{equation}  \label{eq:ej}
E_{J_{\perp }}=z|J_{\perp }|m^{2}-\frac{7}{10}(\pi J_{A}^{\prime
})^{-1/3}(z|J_{\perp }|m)^{4/3} \;.
\end{equation}
Minimizing with respect to $m$ one obtains the equilibrium value of the
sublattice magnetization
\begin{equation}  \label{eq:m}
m=\left( \frac{14}{15}\right)^{3/2} \left( \frac{z|J_{\perp }|}{\pi
J_{A}^{\prime }}\right)^{1/2}\;. 
\end{equation}

The resulting phase diagram is shown in Fig.~\ref{fig:res}. Due to
electron-hole symmetry, the boundaries between the phases do not
depend on the sign of $\Delta$. The spin AF phase AFI, and the spin FM
and orbital AF (FM-AFO) phase dominate the phase diagram. Comparing
Eqs.~(\ref{eq:eaf}) and (\ref{eq:ef}), one obtains a critical value
$U_{c}^{\prime } \simeq 8.52J$ for the boundary between the AFII phase
and the FM-AFO in the limit $|\Delta| \to \infty $.  For $U^{\prime
}>U_{c}^{\prime }$, the only stable phases are the spin AF ones. The
main difference between these phases is that the staggered
magnetization is very small in the AFII phase. In fact, from
Eq.~(\ref{eq:m}) we obtain $m<0.03$ for the parameters of
Fig.~\ref{fig:res}.

Within our perturbation theory up to second order in $t$ and
$t^{\prime }$, the boundary between the FM-AFO phase and the AFII
phase is independent of 
$t$ and $t^{\prime }$. The boundary of the AFI
phase is also weakly dependent on hopping. 
For $U'/J > 2$, the stability region of the AFI phase is slightly
enlarged by increasing the hopping parameters, as can be seen in
Fig~\ref{fig:res}.  For smaller values of $U'/J$ the energy of the
FM-AFO phase decreases due to the proximity of a charge instability
near which our perturbative treatment becomes invalid.  The main
effect of increasing $t$ and $t'$ is to enhance the energy difference
between the stable and unstable phases in each region. These
differences tend to be very small when the hopping parameters are
small. For instance, the energies of the AFII and FM-AFO phases
completely coincide in the limit $t=t'\to 0$.  Thus, narrow-band
systems are likely to show phase coexistence due to inhomogeneities.
Also, we cannot rule out the appearance of more complex phases in a
small region of parameters for which $E_{\text{FM-AFO}} \sim
E_{\text{AFII}}$. One candidate is a phase in which orbitals display
FM order in one direction (say $x$) and AF order in the $y$ direction,
while the spins are ordered antiferromagnetically in the $x$ direction
and ferromagnetically in the $y$ direction.

\section{The doped system}

RuO$_{2}$ planes in RuSr$_{2}$(Eu,Gd)Cu$_{2}$O$_{8}$ are expected to have
electron doping corresponding in our effective Hamiltonian $H$ to a fraction
below 20\% of Ru sites with 4 electrons. Depending on the ratio $\Delta /J$,
there are three possibilities for the ground state of the local Hamiltonian
$H_{I}^{i}+H_{CF}^{i}$ for four electrons (see Table I):

1) $\Delta <0$ --- The ground state is the spin triplet $|i4a_{2}1M\rangle $
(e.g., $|i4a_{2}11\rangle =$ $d_{ixy\uparrow }^{\dagger
}d_{ixy\downarrow }^{\dagger }d_{iyz\uparrow }^{\dagger }d_{izx\uparrow
}^{\dagger }|0\rangle $).

2) $0<\Delta <\Delta _{c}=(\sqrt{41}-1)J/2\simeq 2.70J$ --- The ground state
is the spin triplet and orbital doublet $|i4x1M\rangle $, $|i4y1M\rangle $.

3) $\Delta >\Delta _{c}$ --- The ground state is the spin singlet and orbital
doublet $|i4x0\rangle $, $|i4y0 \rangle $.

Treating the hopping term in first-order degenerate perturbation theory, and
combining with the results of Section IV, we can construct effective
Hamiltonians $H_{\text{eff}}$ for the doped case. We begin by considering
$|\Delta /J|<\sqrt{15}\simeq 3.87$, as suggested by the observed robust AF
order,\cite{lynn} and the results of the previous Section. Then the ground
state of $H_{I}^{i}+H_{CF}^{i}$ for three electrons is the spin quadruplet
with symmetry $B_{1g}$.

\subsection{$ - \sqrt{15} J < \Delta <0$}

In this case, the problem of finding $H_{\text{eff}}$ reduces to calculating
matrix elements of $d_{ixy\sigma }^{\dagger}d_{ixy\sigma }$ (all others
vanish by symmetry) in the basis of $|i3b_{1}\frac{3}{2}M_{1}\rangle $ and
$|i4a_{2}1M_{0}\rangle $. For brevity we shall denote these states as $|i
\frac{3}{2}M_{1}\rangle $ and $|i1M_{0}\rangle $. Using the Wigner-Eckart
theorem, all matrix elements can be calculated in terms of one of them
(e.g., that for maximum projections, which is easily calculated), and
Clebsch-Gordan coefficients $\langle J_{0}jM_{0}m|J_{1}M_{1}\rangle $ for
the combination of angular momenta $J_{0}$ and $j$ to give $J_{1}$. A
similar approach was used before in problems of valence fluctuation
with two magnetic configurations.\cite {bethe} Including the second order
terms described before [Eq.~(\ref{eq:hea})], $H_{\text{eff}}$ becomes
\begin{eqnarray} \label{eq:he1}
H_{\text{eff}}^{(1)} &=& -t^{\prime } \sum_{i\delta \{M\}}\langle
1\frac{1}{2}M_{0}\sigma |\frac{3}{2}M_{1}\rangle 
\langle 1\frac{1}{2}M_{0}^{\prime }\sigma |\frac{3}{2}M_{1}^{\prime }\rangle
\nonumber \\
&&\times |(i+\delta) \frac{3}{2}M_{1}^{\prime }\rangle \langle (i+\delta)
1M_{0}^{\prime }||i1M_{0}\rangle \langle i\frac{3}{2}M_{1}|  \nonumber \\
&&+K\sum_{\langle ij\rangle }({\bf S}_{i}{\bf .S}_{j}-\frac{9}{4})\;,
\end{eqnarray}
where $\{M\}$ denotes the set $M_0,M_1,M_0',M_1'$. Using the same method as
above, it can be easily shown that this model is equivalent to a
double-exchange model with infinite {\em antiferromagnetic} coupling
$J_{\text{de}}$ between localized and itinerant electrons:
\begin{eqnarray} \label{eq:he1b} 
H_{\text{eff}}^{\text{de}} &=& -t^{\prime }\sum_{\langle ij\rangle \sigma
}\left( c_{i\sigma }^{\dag }c_{j\sigma }+{\rm H.c.}\right) +J_{\text{de}
}\sum_{i}{\bf s}_{i}{\bf .S}_{i} \nonumber \\ &&
\mbox{} + K\sum_{\langle ij\rangle }({\bf S}_{i}{\bf
  .S}_{j}-\frac{9}{4})\;.  
\end{eqnarray}
Here $c_{i\sigma }^{\dag }$ is the operator creating an itinerant electron
of spin $\sigma $ at site $i$, and ${\bf s}_{i}=\sum_{\alpha \beta
}c_{i\alpha }^{\dag }\sigma _{\alpha \beta }c_{i\beta }$ gives the spin of
this electron.

The physics of this model is expected to be quite similar to that of the
model with FM exchange, as long as both Ru$^{+5}$ and Ru$^{+4}$ ground-state
configurations are magnetic, which is this case. In fact, treating the spins
classically as in Ref.~\onlinecite{alia}, the sign of 
$J_{\text{de}}$ is irrelevant for the electron dynamics, and only
affects the effective magnetic moment of Ru$^{+4}$. Thus, this results
bring support to the model which successfully explained the magnetic
properties of RuSr$_{2}$(Eu,Gd)Cu$_{2}$O$_{8}$.\cite {alia}

\subsection{$0 < \Delta <\Delta _{c}$}

The 4d$^{4}$ configuration has orbital degeneracy in addition to spin
degeneracy. Proceeding as before, $H_{\text{eff}}$ can again be written as a
double-exchange model, but now there are two types of carriers, each 
hopping only in one direction:
\begin{eqnarray}  \label{eq:he2}
H_{\text{eff}}^{(2)} &=&-t\sum_{i\sigma }(x_{i+{\bf \hat{x}},\sigma }^{\dagger
}x_{i\sigma }+y_{i+{\bf \hat{y}},\sigma }^{\dagger }y_{i\sigma }+\text{H.c.})
\nonumber \\
&&+J^{\prime }\sum_{i}{\bf s}_{i}{\bf .S}_{i}+K\sum_{\langle ij\rangle
}({\bf S}_{i}{\bf .S}_{j}-\frac{9}{4})\;.  
\end{eqnarray}

While the magnetic properties of $H_{\text{eff}}^{(2)}$ should display
some similarities to those of the previous
$H_{\text{eff}}^{\text{de}}$, anisotropic properties and the formation
of stripes are more clearly expected here.  No evidence of stripes in
this system has been reported so far. A preferential direction was not
observed in neutron experiments. \cite{lynn} However, an equal amount
of small domains with stripes oriented in the $x$ and $y$ direction
cannot be completely ruled out by these experiments.
Furthermore, although we explore here, for completeness, all the
possibilities of our model, the available experimental results seem to
indicate that $\Delta < 0$.

\subsection{$\Delta _{c} < \Delta < \sqrt{15} J$}

In this case, the first order effective hopping vanishes. It is
necessary to go to third order in $H_{h}$ to effectively exchange a
4d$^{3}$ spin quadruplet with a 4d$^{4}$ spin singlet and orbital
doublet. Thus, the added electrons are essentially localized, and the
observed magnetic properties would be difficult to explain within this
picture.

\subsection{$|\Delta| > \sqrt{15} J$}

Now, for any sign of $\Delta $, the ground state of $H_{I}^{i} + H_{CF}^{i}$
for $n=3$ is the spin and orbital doublet $|i3x\frac{1}{2} M\rangle $,
$|i3y\frac{1}{2}M\rangle $. We assume that the system is in the AFII phase
to be consistent with neutron experiments, in spite of the different
magnitude of the localized moment. Thus, the orbital degree of freedom is
ferromagnetically frozen in the direction $\gamma =x$ or $y$. For negative
$\Delta $, $H_{\text{eff}}$ turns out to be equivalent to a double-exchange
model with itinerant electrons coupled ferromagnetically to the localized
spins 1/2, and one-dimensional hopping in the direction {\em perpendicular}
to $\gamma $. Instead, for positive $\Delta > \sqrt{15} J$, the resulting
$H_{\text{eff}}$ is equivalent to a $t\,$--$\,J$ model with isotropic
hopping and anisotropic exchange.

\section{Discussion}

We have studied the electronic structure of RuO$_{2}$ planes in
RuSr$_{2}$(Eu,Gd)Cu$_{2}$O$_{8}$ using a strong-coupling approach to
describe the 4d $t_{2g}$ orbitals of Ru and their interactions. For
undoped planes (corresponding to formal valence +5 for Ru ions), we
find three possible phases. Two of them are favored for large
tetragonal crystal field (of any sign), and have orbital degrees of
freedom which order at zero temperature (also at finite temperatures
if hopping along the tetragonal axis were included). The spins order
either ferromagnetically or in a particular AF order with very small
staggered magnetic moment compared to the experimentally observed one
$m\sim 1.2\mu _{B}$,\cite{lynn} due to strong one-dimensional
fluctuations.  The dominant phase for small crystal-field splitting
consists of spins 3/2 which order antiferromagnetically, with
nearest-neighboring spins pointing in opposite directions, as observed
in neutron experiments.\cite{lynn} One might wonder whether the
effective measured staggered moment $\sim 1.2\mu _{B}$ is closer to
that of a localized spin 1/2 rather than 3/2. However, for both
mentioned AF phases there are several physical ingredients that {\em
reduce} the measured moment: i) spin fluctuations that reduce the
sublattice magnetization, ii) effective Ru-Ru charge fluctuations
(that are easily calculated within our perturbative approach), iii)
Ru-O charge fluctuations,\cite{pet} and iv) doping, particularly if
ferromagnetic polarons are formed.\cite{alia}

When the system is doped with electrons, there are two main
possibilities depending on the sign of the tetragonal crystal field
parameter $\Delta $.  If it is negative (as it seems to be the
case\cite{lieb} in Sr$_{2}$RuO$_{4}$), the additional carriers are
described by a double-exchange model with infinite AF coupling with
the localized $S=3/2$ spins. This model is able to qualitatively
explain the apparent contradiction between observed antiferromagnetic
order, magnetic field dependence of the magnetization, and temperature
dependence of the magnetic susceptibility.\cite{alia} It is also
consistent with the observed magnetoresistance.\cite{awan,chen}.
Using previous results of the effective double exchange model
\cite{alia}, the experimental positive Curie constant $\Theta =100\pm
3$ K suggests that the $xy$ hopping $t^{\prime }\sim 0.25$ eV.  A more
quantitative description of the magnetic properties requires an
accurate calculation of the magnetic moment.  It is also possible that
the double-exchange model should be supplemented by inter-atomic
Coulomb repulsions of a moderate range,\cite{garc} since the number of
carriers in the system is low, particularly taking into account the
low superconducting critical temperature. Also, previous studies of
the double-exchange model suggest that there is macroscopic phase
separation at small doping\cite{alia2,sev} which is inhibited by
long-range Coulomb repulsion.

If $\Delta$ is positive, the effective model 
for the doped case is similar, but the carriers have an orbital degree
of freedom, as in manganites,\cite{ishih,fein} which might lead to the
observation of orbitons by Raman scattering\cite{sait} for enough
doping. We also expect the formation of stripes in this case. 
However, there is no experimental evidence of stripes so far in the
system, and fitting of optical properties of another layered ruthenate
Sr$_{2}$RuO$_{4}$ suggests that $\Delta$ is small and
negative.\cite{lieb} 
More detailed studies of the effects of doping would be possible if
the superconducting critical temperature $T_{S}$ could be further
enhanced, either by appropriate substitution of the rare earth or by
applied pressure.

\acknowledgements

One of us (AAA) wants to thank A. Trumper, B. Normand, and G. Mart\'{i}nez
for useful discussions. AAA partially supported by CONICET. This work was
sponsored by PICT 03-12742 of ANPCyT. We acknowledge support from CAPES,
a division of the Brazilian Ministry of Education.

\end{document}